\documentstyle[11pt,epsfig,amsbsy]{article}

\def\be{\begin{equation}}
\def\ee{\end{equation}}
\def\ba{\begin{eqnarray}}
\def\ea{\end{eqnarray}}

\def\ga{\mathrel{\raise.3ex\hbox{$>$\kern-.75em\lower1ex\hbox{$\sim$}}}}
\def\la{\mathrel{\raise.3ex\hbox{$<$\kern-.75em\lower1ex\hbox{$\sim$}}}}

\newcommand{\fr}[2]{\frac{#1}{#2}}

\newcommand{\DE}{\rm{DE}}
\begin{document}

\baselineskip=16pt
\begin{titlepage}
\begin{center}

\vspace{0.5cm}

\large {\bf Lee Wick Dark Energy}
\vspace*{5mm} \normalsize

{\bf Seokcheon Lee}

\smallskip
\medskip

{\it Institute of Physics, Academia Sinica, \\
Taipei, Taiwan 11529, R.O.C.}

\smallskip
\end{center}

\vskip0.6in

\centerline{\large\bf Abstract}

We consider the cosmological application of Lee-Wick theory where a field has a higher derivative kinetic operators. The higher derivative term can be eliminated by introducing a set of auxiliary fields. We investigate the cosmological evolutions of these fields as a candidate of dark energy. This model has the same structure as so called ``quintom' model except the form of potentials and the sign of the slope of the potentials. This model can give the stable late time phantom dominated scaling solution ($\omega_{\DE} < -1$) or tracking attractors ($\omega_{\DE} = 0$) depending on the choice of the slopes of the potential. In order to be a viable dark energy candidate, the present energy density contrast of dark energy  ($\Omega_{\DE}^{(0)}$) should be close to an observed value ($0.73$) at the same time. However, a simple toy model of the theory can not satisfy both $\omega_{\DE}^{(0)} \simeq -1$ and $\Omega_{\DE}^{(0)} = 0.73$. If we include the self interaction term of Lee Wick field, then we are able to produce the observed values of those quantities. However, in this case we are not be able to have the stable solutions and we need to suffer from the fine tuning of its mass. 

\vspace*{2mm}

\end{titlepage}

\section{Introduction}
\setcounter{equation}{0}

The instability of the Higgs potential due to the quadratic radiative corrections has motivated many extensions beyond the Standard Model (SM). Recently, there is a new extension of the SM so called Lee-Wick Standard Model (LWSM) \cite{Grinstein}, based on the work of Lee and Wick \cite{Lee, Wick} where they show a finite theory of quantum electrodynamics (QED) by including the regulator term as a higher derivative version of QED. If we extend the idea to all the SM field with a higher derivative term, then we have a LW partner for the every SM field.  Its application to gauge bosons \cite{07043458}, neutrino masses \cite{07051188}, Maxwell-Einstein theory \cite{07053287}, flavor changing \cite{07080567}, and electroweak precision data \cite{08021061, 08053296, 08064555} have been considered.

There is an almost identical work with the above proposal to demonstrate a single scalar field model of dark energy with a higher derivative shown the equation of state (eos) crossing $-1$ \cite{0503268, 0603824}. Similar idea adapted to inflation model is also considered \cite{0504560}. By including the only self interaction term of the phantom field ($M^2 \psi^2/2$), this model can provide $\Omega_{\DE}^{(0)} = 0.73$ and $\omega_{\DE}^{(0)} \simeq -1.0$. However, this kind of potential suffers from the fine tuning problem. Including the higher derivatives of a scalar field is well match to one of dark energy models, so called ``Quintom'' model \cite{0404062} when we introduce a set of auxiliary fields. There have been many works done for quintom models with various potentials \cite{0410654, 0412308, 0501652, 0602590, 0604460, 0608165, 0701353}. The perturbation of quintom models and their effects on cosmological observations are also studied \cite{0507482, 07112187}. Even though there are potential problems of this model like the violation of causality and the instability of ghost field, we will not consider these aspects in this work.

The energy density of the Lee Wick (LW) field dominates at late time because its energy density increases ($\omega_{\psi} < -1$) while those of the scalar field and the barotropic fluid decrease ($\omega_{\phi}, \omega_{\gamma} > -1$) as the universe evolves. However, the eos of the effective dark energy ({\it i.e.} the combination of $\omega_{\phi}$ and $\omega_{\psi}$) can cross $-1$ depending on the slopes of potentials of two fields. This dark energy might be able to preserve the tracking behavior of the scalar field requiring less fine tuning at early universe and reach to phantom attractors at late time. Only self interaction of two fields are considered in many literatures except \cite{0501652, 0602590}. However, we naturally have the interaction between the scalar field and the LW field in LW theory.

In the next section, we will investigate the existence and the stability of tracker solutions of LW dark energy for the specific choice of potential by using the analysis of autonomous system for physical quantities. We check the cosmological evolutions of the dark energy and the barotropic fluid in section 3. We reach our conclusions in section 4.

\section{Lee-Wick Dark Energy}
\setcounter{equation}{0}

To illustrate the effect of Lee-Wick theory in cosmology, we consider a theory of one self interacting scalar field, $\sigma$, with a higher derivative term. The Lagrangian density is

\be {\cal L}_{\sigma} = \fr{1}{2} \nabla_{\mu} \sigma \nabla^{\mu} \sigma - \fr{1}{2 M^2} (\Box \sigma)^2 - V(\sigma) \,  , \label{Lsigma} \ee where $\Box \sigma = g^{\mu \nu} \nabla_{\mu} \nabla_{\nu} \sigma$. We can rewrite the above Lagrangian by introducing an auxiliary scalar field $\psi$ as

\be {\cal L}_{\sigma - \psi} = \fr{1}{2} \nabla_{\mu} \sigma \nabla^{\mu} \sigma - \psi \Box \sigma + \fr{1}{2} M^2 \psi^2 - V(\sigma) \, , \label{Lsigmapsi} \ee where $\Box \sigma = M^2 \psi$ can be obtained from the above Lagrangian (\ref{Lsigmapsi}) to recover the original Lagrangian (\ref{Lsigma}). Now if we define $\sigma = \phi - \psi$, then the Lagrangian (\ref{Lsigmapsi}) becomes

\be {\cal L}_{\phi-\psi} = \fr{1}{2} \nabla_{\mu} \phi \nabla^{\mu} \phi - \fr{1}{2} \nabla_{\mu} \psi \nabla^{\mu} \psi + \fr{1}{2} M^2 \psi^2 - V(\phi - \psi) \, , \label{Lpsiphi} \ee where we use the integration by parts \cite{0503268, 0603824}. In this form there are two kinds of scalar field, a normal scalar field $\phi$ and an Lee-Wick (LW) field which has the opposite sign of the kinetic term compared to the normal field. Stability of this theory was considered in the literature \cite{Grinstein}. We will present a phase-space analysis of these scalar fields in a homogeneous and isotropic flat universe containing a barotropic fluid with its eos ($\omega_{\gamma} = \gamma - 1$) where the pressure and the energy density of the fluid is related by $p_{\gamma} = (\gamma - 1) \rho_{\gamma}$.

As a toy model, we ignore the mass term in $\psi$ and assume the interacting terms of two scalar fields as
an exponential potential. \be V(\phi-\psi) = V_{\rm{int}}(\phi,\psi) = V_{\phi-\psi}^{0} \exp(-\kappa \lambda_{\phi} \phi - \kappa \lambda_{\psi} \psi) \, , \label{Vphipsi} \ee where $\kappa = \sqrt{8 \pi G}$. As shown in the above Lagrangian (\ref{Lpsiphi}) we can not help having the interaction term between two scalar fields in LW theory.  The homogeneous scalar fields $\phi$ and $\psi$ in a spatially flat FRW cosmological model can be described by the fluids with effective energy density and the pressure

\ba \rho_{\DE} &=& \fr{1}{2} \dot{\phi}^2 -\fr{1}{2} \dot{\psi}^2 + V_{\rm{int}}(\phi,\psi) = \rho_{\phi} + \rho_{\psi} - V_{\rm{int}}(\phi,\psi) \, , \label{rhoDEint} \\
 p_{\DE} &=& \fr{1}{2} \dot{\phi}^2 -\fr{1}{2} \dot{\psi}^2 - V_{\rm{int}}(\phi,\psi) = p_{\phi} + p_{\psi} + V_{\rm{int}}(\phi,\psi) \equiv \omega_{\DE} \rho_{\DE} \, , \label{pDEint} \ea where we indicate the effective eos of the dark energy as $\omega_{\DE}$ and define the energy densities and pressures of a scalar field and a LW field as \ba \rho_{\phi} &=& \fr{\dot{\phi}^2}{2} + V_{\rm{int}}(\phi,\psi) , \hspace{0.2in} p_{\phi} = \fr{\dot{\phi}^2}{2} - V_{\rm{int}}(\phi,\psi) \equiv \omega_{\phi} \rho_{\phi} \, , \label{rhopphi} \\ \rho_{\psi} &=& -\fr{\dot{\psi}^2}{2} + V_{\rm{int}}(\phi,\psi) , \hspace{0.2in} p_{\psi} = - \fr{\dot{\psi}^2}{2} - V_{\rm{int}}(\phi,\psi) \equiv \omega_{\psi} \rho_{\psi} \, . \label{rhoppsi} \ea We show the second equality of dark energy density and pressure related to those of a scalar field and a LW field because we have the interaction between two fields and only the total energy-momentum will be conserved.
Now the Fridemann equation becomes
\be H^2 = \fr{\kappa^2}{3} \Biggl(\rho_{\DE} + \rho_{\gamma} \Biggr) = \fr{\kappa^2}{3} \Biggl(\rho_{\phi} + \rho_{\psi} - V_{\rm{int}}(\phi,\psi) + \rho_{\gamma} \Biggr) \equiv \fr{\kappa^2}{3} \rho_{cr} \, , \label{H2int} \ee where $\rho_{cr}$ is the critical energy density and we again explicitly represent $\rho_{\DE}$ in the second equality to show the differences in the evolution equations with and without the interaction term between two scalar fields. Then the effective eos of the dark energy $\omega_{\DE}$ is given by
\be \omega_{\DE} = \fr{p_{\DE}}{\rho_{\DE}} = \fr{\dot{\phi}^2 - \dot{\psi}^2 - 2V_{\rm{int}}(\phi,\psi)}{\dot{\phi}^2 - \dot{\psi}^2 + 2V_{\rm{int}}(\phi,\psi)} \, .\label{omegaDEint} \ee This effective eos can be rewritten by using equations (\ref{rhopphi}) and (\ref{rhoppsi}) as \be \omega_{\DE} = \fr{1}{3} \Biggl(1 +  \fr{2(\omega_{\phi} \Omega_{\phi} + \omega_{\psi} \Omega_{\psi})}{\Omega_{\DE}} \Biggr) \, , \label{omegaDEint2} \ee where $\Omega_{\DE} = \fr{\rho_{\DE}}{\rho_{cr}}$. We can find the evolution equations of two scalar fields and the fluid in flat FRW models by using the Bianchi identity which are

\ba \ddot{\phi} + 3H \dot{\phi} + \fr{\partial V_{\rm{int}}(\phi,\psi)}{\partial \phi} &=& 0 \, , \label{dotphiint} \\ \ddot{\psi} + 3H \dot{\psi} - \fr{\partial V_{\rm{int}}(\phi, \psi)}{\partial \psi} &=& 0 \, ,\label{dotpsiint} \\ \dot{\rho}_{\gamma} + 3H \gamma \rho_{\gamma} &=& 0 \, .\label{dotrhogammaint} \ea We can find that the energy density of an individual scalar field is not conserved while the total energy density is conserved \ba \dot{\rho}_{\phi} + 3H (1 + \omega_{\phi}) \rho_{\phi} &=& \fr{\partial V_{int}(\phi,\psi)}{\partial \psi} \dot{\psi} \, ,\label{dotrhophi} \\ \dot{\rho}_{\psi} + 3H (1 + \omega_{\psi}) \rho_{\psi} &=& \fr{\partial V_{int}(\phi,\psi)}{\partial \phi} \dot{\phi} \, , \label{dotrhopsi}  \\ \dot{\rho}_{\DE} + 3H (1 + \omega_{\DE}) \rho_{\DE} &=& 0 \, . \label{dotrhoDE} \ea
To study the critical point structure and the stability of the system, it is convenient to introduce the following dimensionless variables \cite{Copeland}

\ba x_{\phi} &=& \fr{\kappa \phi'}{\sqrt{6}} \, , \hspace{0.2in} y = \fr{\kappa \sqrt{V_{\rm{int}}(\phi,\psi)}}{\sqrt{3} H} \, , \label{xyphiint} \\ x_{\psi} &=& \fr{\kappa \psi'}{\sqrt{6}} \, , \hspace{0.2in} z = \fr{\kappa \sqrt{\rho_{\gamma}}}{\sqrt{3} H} \, . \label{zint} \ea where a prime means a derivative with respect to the logarithm of the scale factor $\ln a$. With these variables (\ref{xyphiint} - \ref{zint}) we can express the Fridemann equation (\ref{H2int}) and its derivative as

\ba 1 &=& x_{\phi}^2 - x_{\psi}^2 + y^2 + z^2 \, , \label{H3int} \\ \fr{H'}{H} &=& - 3 \Bigl( x_{\phi}^2 - x_{\psi}^2 + \fr{\gamma}{2} z^2 \Bigr) = 3 \Biggl[ -\fr{\gamma}{2} + \Bigl(\fr{\gamma}{2} -1 \Bigr) x_{\phi}^2 - \Bigl(\fr{\gamma}{2} -1 \Bigr) x_{\psi}^2 + \fr{\gamma}{2} y^2 \Biggr] \, .\label{Hprimeint} \ea Also we can find the evolution equation of those variables (\ref{xyphiint} - \ref{zint})

\ba x_{\phi}' &=& - \Biggl( \fr{H'}{H} + 3 \Biggr) x_{\phi} + \fr{\sqrt{6}}{2} \lambda_{\phi} y^2 \, ,\label{xphipint} \\ y' &=& - y \Biggl( \fr{H'}{H} + \fr{\sqrt{6}}{2} ( \lambda_{\phi} x_{\phi} + \lambda_{\psi} x_{\psi}) \Biggr) \, ,\label{ypint} \\ x_{\psi}' &=& - \Biggl( \fr{H'}{H} + 3 \Biggr) x_{\psi} - \fr{\sqrt{6}}{2} \lambda_{\psi} y^2 \, .\label{xpsipint} \ea Fixed points at finite values of $x_{\phi}, x_{\psi}$ and $y$ in the phase space are correspond to solutions where the scalar field(s) has a constant equation of state. Critical points correspond to fixed points where $x_{\phi}' = y' = x_{\psi}' = 0$ and give an expanding universe with a scale factor given by $a \propto t^{\alpha}$ where \be \alpha = \fr{2}{6 x_{\phi}^2 - 6 x_{\psi}^2 + 3 \gamma z^2} \, .\label{alphaint} \ee

\begin{table*}
\begin{center}
\caption{The values and the existence condition for fixed points. The various tracking attractor solutions are shown for the specific choices of $\lambda_{\phi}$ and $\lambda_{\psi}$. We define $ \lambda_{eff}^2 = \lambda_{\phi}^2-\lambda_{\psi}^2$.
\label{tab:T1int}}
\vskip .3cm
\small
\begin{tabular}{cccccccc}
\hline Case & $ x_{\phi} $ & $y$  & $x_{\psi}$ & $z$ &
Existence & $\Omega_{\DE}$ & $\gamma_{\DE}$ \\
\hline
$K$ & $ x_{\phi}$ & $0$ &
$\sqrt{x_{\phi}^2 -1}$ &
$0$ & All $\lambda_{\phi}, \lambda_{\psi}$ & $1$ &$2$ \\
$F$ & $0$ & $0$ & $0$ & $1$& All $\lambda_{\phi}, \lambda_{\psi}$ &$0$ &
undefined \\

$SP$ & $\fr{\lambda_{\phi}}{\sqrt{6}}$ & $\sqrt{1-\fr{\lambda_{eff}^2}{6}}$ &
$\fr{\lambda_{\psi}}{\sqrt{6}}$ & $0$& $\lambda_{eff}^2 < 6$ & $1$ & $\fr{\lambda_{eff}^2}{3}$ \\
$SPF$ & $\fr{\sqrt{6}\gamma}{2}\fr{\lambda_{\phi}}{\lambda_{eff}^2}$ & $\sqrt{\fr{3\gamma(2-\gamma)}{2 \lambda_{eff}^2}}$ &
$-\fr{\sqrt{6}\gamma}{2}\fr{\lambda_{\psi}}{\lambda_{eff}^2}$ &
$\sqrt{1-\fr{3\gamma}{\lambda_{eff}^2}}$ & $ \lambda_{eff}^2 > 3\gamma$ & $\fr{3\gamma}{\lambda_{eff}^2}$ & $\gamma$ \\

\hline
\end{tabular}
\end{center}
\end{table*}

In Table \ref{tab:T1int}, we classify critical points and show their properties and values. We show fixed points as one two dimensional hyperbola $K$ which is kinetic dominated solutions, a fixed point $F$ corresponding to a barotropic fluid dominated solution, a fixed point $SP$ which is dark energy dominated solution, and a fixed point $SPF$ meaning a fluid dark energy dominated solution. $K$ and $F$ points exist for any value of $\lambda_{\phi}$ and $\lambda_{\psi}$. However, there are constraints for the existence of points $SP$ and $SPF$. The point $SP$ exists only when $\lambda_{eff}^2 \equiv \lambda_{\phi}^2 - \lambda_{\psi}^2 < 6$. Thus, if the slope of a potential of a scalar field $\lambda_{\phi}$ is too big compared to that of a LW field $\lambda_{\phi}$, then a point $SP$ will not be existed. In this point the effective equation of state of dark energy is given by $\omega_{\DE} = \fr{\lambda_{eff}^2}{3} -1$. Also the point $SPF$ exists only for $\lambda_{eff}^2 > 3 \gamma$ cases. In this case, a scalar field evolves much faster than a LW field and it tracks the background fluid. Thus, the effective equation of state of dark energy follows that of a barotropic fluid $\omega_{\DE} = \omega_{\gamma}$.

We study the stability of critical points by substituting linear perturbations of variables (\ref{xyphiint}) - (\ref{zint}) into the evolution equations (\ref{Hprimeint}) - (\ref{xpsipint}) and find three eigenvalues for the evolution equations of the first order in the perturbations. Stability of system requires the real part of all eigenvalues to be negative.

\begin{table*}
\begin{center}
\caption{Eigenvalues and stability
\label{tab:T2int}}
\vskip .3cm
\small
\begin{tabular}{cccccc}
\hline Case & $m_1$ & $m_2$  & $m_3$ & Stability \\
\hline
K & $0$ & $3(2-\gamma)$ & $3-\sqrt{\fr{3}{2}}(\lambda_{\phi} x_{\phi} + \lambda_{\psi} x_{\psi})$ & unstable \\
F & $-\fr{3}{2}$ & $-\fr{3}{2}$ & $\fr{3}{2}$ & unstable \\
SP & $-3 + \fr{\lambda_{eff}^2}{2}$ & $-3 + \fr{\lambda_{eff}^2}{2}$ &
$\lambda_{eff}^2-3\gamma$ & stable \\
& & & & if $\lambda_{eff}^2 < 3 \gamma$ \\
SPF & $-\fr{3}{2}(2-\gamma)$ & $-\fr{3}{4}(2-\gamma)\Biggl[1+\sqrt{1+\fr{8\gamma(3\gamma-\lambda_{eff}^2)}{(2-\gamma)\lambda_{eff}^2}}\Biggr]$ &
$-\fr{3}{4}(2-\gamma)\Biggl[1-\sqrt{1+\fr{8\gamma(3\gamma-\lambda_{eff}^2)}{(2-\gamma)\lambda_{eff}^2}}\Biggr]$ &
 stable \\
& & & & if $3\gamma < \lambda_{eff}^2 < \fr{24 \gamma^2}{9\gamma-2}$\\

\hline
\end{tabular}
\end{center}
\end{table*}

$K$ : These kinetic dominated solutions always exist for any value of $\lambda_{\phi}$ and $\lambda_{\psi}$. It gives the equation of state of dark energy as $1$ like inflation. This yields $H'/H = -3$. The linearized system of (\ref{Hprimeint}) - (\ref{xpsipint}) about these fixed points gives three eigenvalues \be 0, \hspace{0.2in} 3(2-\gamma), \hspace{0.2in} 3-\sqrt{\fr{3}{2}}(\lambda_{\phi} x_{\phi} + \lambda_{\psi} x_{\psi}) \, . \label{keigenint} \ee
Thus the kinetic dominated solutions are unstable.

$F$ : The fluid dominated solution also exists for any $\lambda_{\phi}$ and $\lambda_{\psi}$. There is no dark energy contribution. The eigenvalues for the system are \be -\fr{3}{2}, \hspace{0.2in} -\fr{3}{2}, \hspace{0.2in} \fr{3}{2} \, , \label{feigenint} \ee which show that the solution is unstable.

$SP$ : The dark energy dominated solution exists for $\lambda_{\phi} < \sqrt{\lambda_{\psi}^2 + 6}$. This means that as long as the slope of the potential of a scalar field $\lambda_{\phi}$ is not much stiffer than that of LW field $\lambda_{\psi}$ we can have the dark energy dominated solution. The eigenvalues for the system at this point are \be -3 + \fr{\lambda_{eff}^2}{2}, \hspace{0.2in} -3 + \fr{\lambda_{eff}^2}{2}, \hspace{0.2in}  \lambda_{eff}^2-3\gamma \, , \label{speigenint} \ee which indicate that the solution is stable when the constraint $\lambda_{eff}^2 < 3\gamma$ is satisfied. Unlike a single scalar field models, one can obtain the accelerated expansion even if the potential is not too flat. To obtain the accelerating expansion we need that the slope of a potential of a scalar field should not be too much flatter than that of a LW field. Similar idea is adapted in so called ``assisted quintessece'' \cite{Coley, SKim}.

$SPF$ : The dark energy fluid dominated solution exits for $\lambda_{eff}^2 > 3\gamma$. This solution tracks the background barotropic fluid. Thus we need a stiff scalar field potential compared to that of LW field. At this point, the eigenvalues for the system are \be -\fr{3}{2}(2-\gamma), \hspace{0.2in}  -\fr{3}{4}(2-\gamma)\Biggl[1+\sqrt{1+\fr{8\gamma(3\gamma-\lambda_{eff}^2)}{(2-\gamma)\lambda_{eff}^2}}\Biggr], \hspace{0.2in}
-\fr{3}{4}(2-\gamma)\Biggl[1-\sqrt{1+\fr{8\gamma(3\gamma-\lambda_{eff}^2)}{(2-\gamma)\lambda_{eff}^2}}\Biggr] \, . \label{spfeigenint} \ee Thus, we have the stable solution as long as the constraint $3\gamma< \lambda_{eff}^2 < 24\gamma^2/(9\gamma-2)$ is satisfied.

\section{Evolution without Self Interaction}
\setcounter{equation}{0}

In the previous section, we study the critical points and their existences and stabilities. The effective eos of the dark energy is given by $\omega_{\DE} = -1 + \lambda_{eff}^2/3$ at the point $SP$. Thus, we should have $\lambda_{eff}^2 < 2$ in order to get the accelerating expansion at late time ({\it i.e.} $\omega_{\DE} < -1/3$) when the dark energy dominates the Universe. We can see the situation more clearly if we express the effective dark energy energy contrast $\Omega_{\DE}$ and its eos $\omega_{\DE}$ by using the dimensionless variables in (\ref{xyphiint}) and (\ref{zint}) \ba \Omega_{\DE} &=& x_{\phi}^2 - x_{\psi}^2 + y^2 \, , \label{OmegaDEint} \\ \omega_{\DE} &=& \fr{x_{\phi}^2 - x_{\psi}^2 - y^2}{x_{\phi}^2 - x_{\psi}^2 + y^2} \, .\label{omegaDEint3} \ea From these equation, we can find the condition for late time accelerating universe. Energy density of dark energy component should be dominant and its eos should be smaller than $-1/3$ to get the late time accelerated expansion. In Table \ref{tab:T3}, we distinguish a scalar field dominated case ($\omega_{\DE}^{(0)} > -1$) with a LW field dominated cases ($\omega_{\DE}^{(0)} < -1$) for the specific choices of $\lambda_{\phi}$ and $\lambda_{\psi}$. We mean the present values of any quantities by using $(0)$ in the affix. For example, we obtain $x_{\phi}^{(0)} > x_{\psi}^{(0)}$ from the above equation (\ref{omegaDEint3}) to get a scalar field dominated universe at present. This condition can be satisfied if the slope of the potential of a scalar field $|\lambda_{\phi}|$ is stiffer than that of a LW field $|\lambda_{\psi}|$. In this case a scalar field rolls a potential faster than a LW field and satisfy the above condition. Late time phantom dominated epoch $\omega_{\DE} < -1$ also can be reached when a LW field moves faster than a scalar field. This case will be happened when $|\lambda_{\psi}| > |\lambda_{\phi}|$. Only the magnitude of slope is concerned in those cases. We have the opposite signs of the slopes for scalar fields in LW theory as shown in the equation (\ref{Lpsiphi}).  When $|\lambda_{\phi}| = |\lambda_{\psi}|$, the scalar field $\phi$ and a LW field $\psi$ evolves same as each other. In this case, we have $x_{\phi}^2 = x_{\psi}^2$ and the eos of dark energy is always satisfied $\omega_{\DE} = -1$ even though each scalar field evolves.

\begin{table*}
\begin{center}
\caption{The present value of dark energy eos $\omega_{\DE}$ for the various conditions for $\lambda_{\phi}$ and $\lambda_{\psi}$. The slopes of potential determine the values of $x_{\phi}$ and $x_{\psi}$ at present giving the different values of $\omega_{\DE}^{(0)}$. In this table $x_{\phi}$ and $x_{\psi}$ represent the present values of them.
\label{tab:T3}}
\vskip .3cm
\small
\begin{tabular}{ccccc}
\hline $\omega_{\DE}^{(0)}$ & $0$ &  $ < -1$ & $-1$  & $> -1$ \\
\hline
$|\lambda_{\phi}|$ & $> \sqrt{\lambda_{\psi}^2 + 3\gamma}$ & $< |\lambda_{\psi}|$ & $|\lambda_{\psi}|$ & $ > |\lambda_{\psi}|$ \\
$x_{\phi}^{2}$ & $ x_{\psi}^2 + y^2$ & $ < x_{\psi}^2$ & $x_{\psi}^2$ & $ > x_{\psi}^2$  \\
\hline
\end{tabular}
\end{center}
\end{table*}

We show the evolutions of energy density contrast of dark energy, that of the matter, and the eos of dark energy for the specific values of $\lambda_{\phi}$ and $\lambda_{\psi}$.

\begin{center}
\begin{figure}
\vspace{1.5cm}
\centerline{
\psfig{file=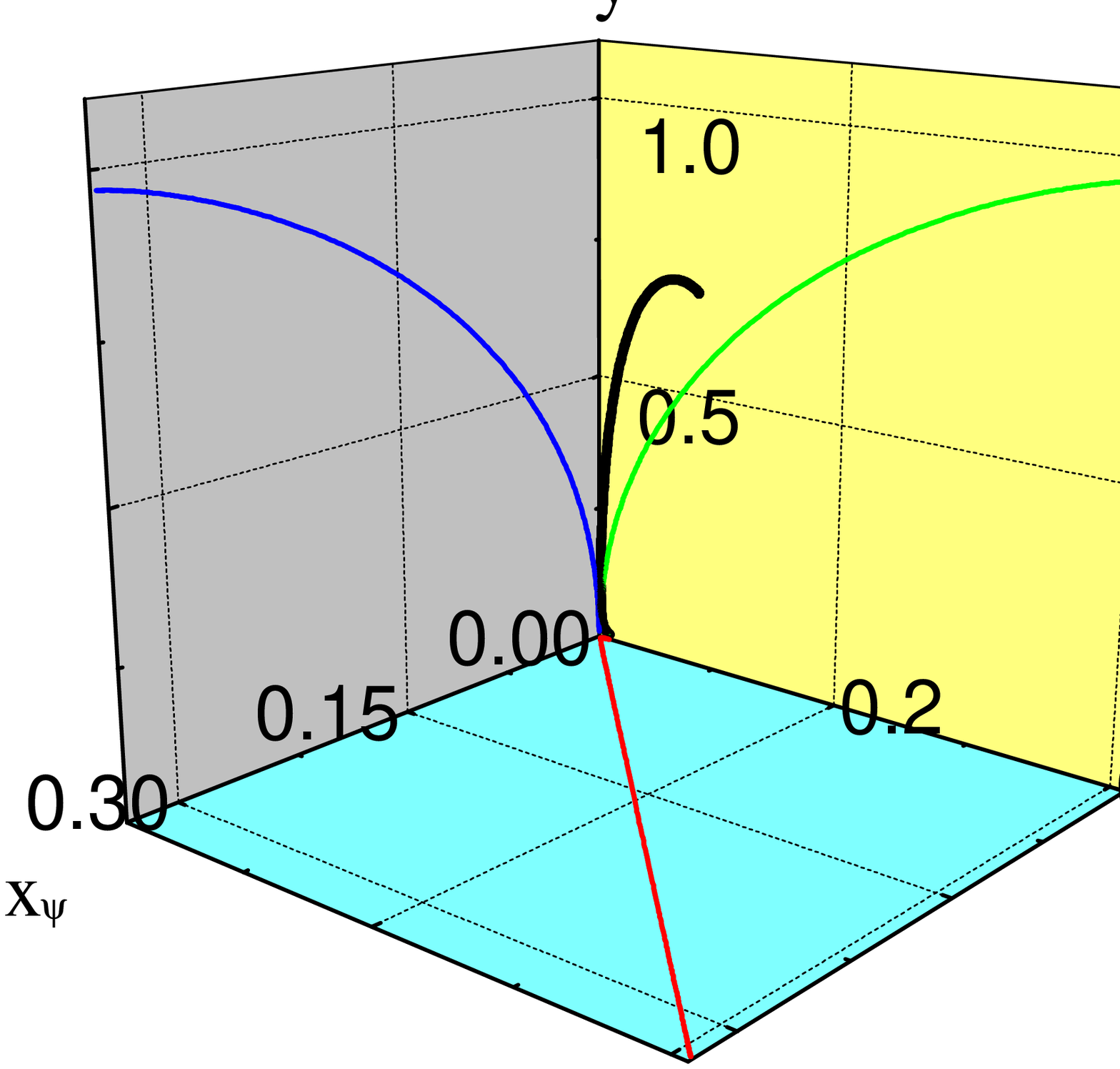, width=8cm}\psfig{file=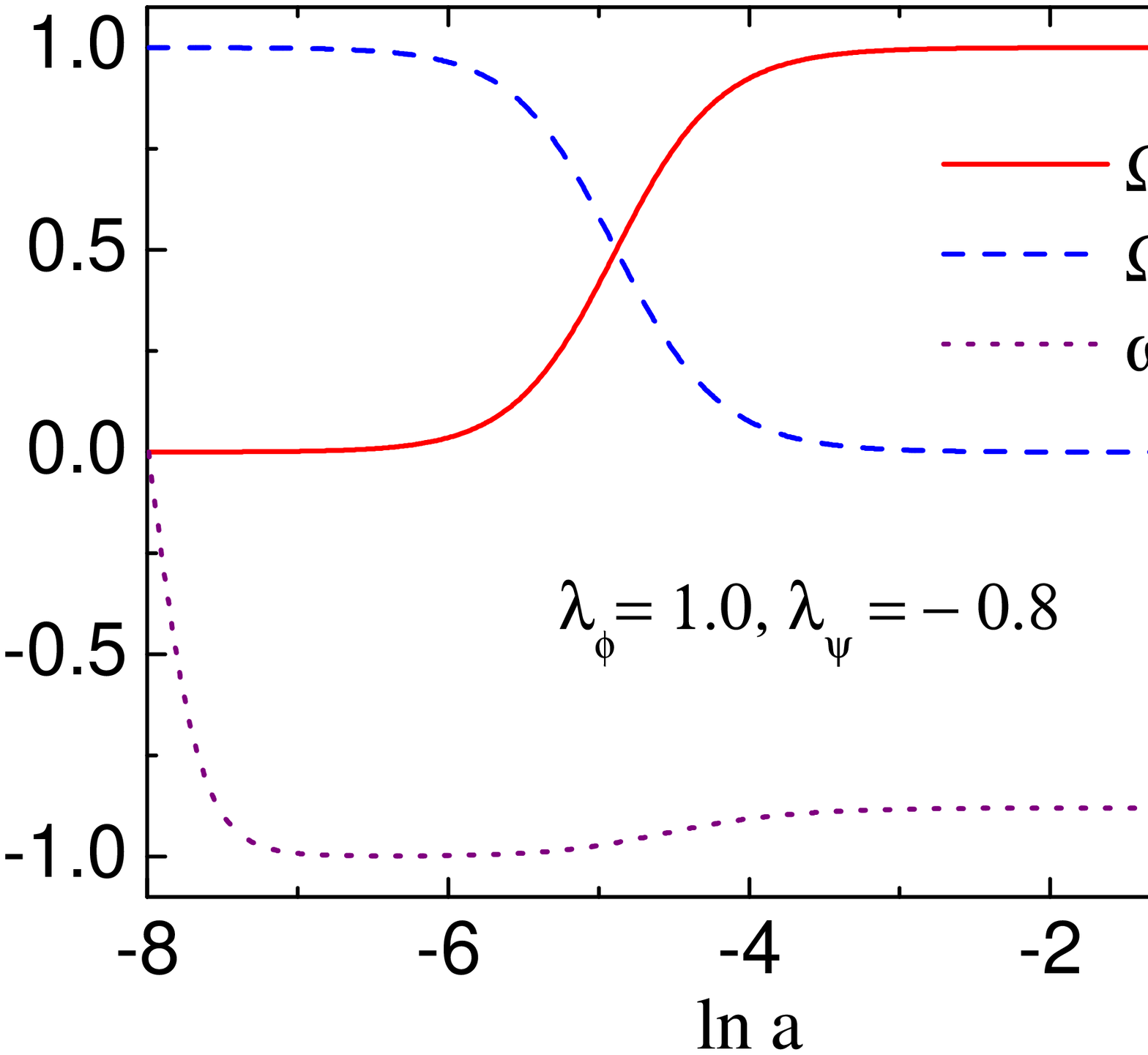, width=8cm} }
\vspace{-1.0cm}
\caption{ a) The evolution of $x_{\phi}$, $x_{\psi}$, and $y$ for $\lambda_{\phi} = 1.0$ and $\lambda_{\psi} = - 0.8$. b) The evolution of $\Omega_{\DE}$, $\Omega_{m}$, and $\omega_{\DE}$ for the same values of $\lambda_{\phi}$ and $\lambda_{\psi}$
.} \label{fig10m08}
\end{figure}
\end{center}

Case I ) $|\lambda_{\phi}| > |\lambda_{\psi}|$ :

In the left panel of Fig. \ref{fig10m08} we show the evolutions of dimensionless variables $x_{\phi}$, $x_{\psi}$, and $y$ when $|\lambda_{\phi}| = 1.0 > |\lambda_{\psi}| = 0.8$. In this case, a scalar field rolls down a potential faster than a LW field. Thus, we have bigger $x_{\phi}$ ({\it i.e.} $\phi'$) value than $x_{\psi}$ ({\it i.e.} $\psi'$) as shown in the left panel of Fig. \ref{fig10m08}. The $y$ value is also increased as fields evolve. We can understand this evolutions from equations (\ref{xphipint}) - (\ref{xpsipint}). All of $x_{\phi}$, $x_{\psi}$, and $y$ values are increased as they evolve in the universe. In the $x_{\phi}$-$x_{\psi}$ plane, we can see that $x_{\phi}$ starts from $0$ and reaches to $0.41$ while $x_{\psi}$ increases from $0$ to $0.33$. In both $x_{\phi}$-$y$ and $x_{\psi}$-$y$ planes, we find that $y$ is increased from $0$ to $0.97$. Because $x_{\phi}$ is bigger than $x_{\psi}$, $y$ is smaller than $1$ as can be seen in the equation (\ref{OmegaDEint}). Also the evolutions of the energy density contrast of dark energy $\Omega_{\DE}$, that of dark matter $\Omega_{m}$, and the eos of the dark energy $\omega_{\DE}$ are shown in the right panel of Fig. \ref{fig10m08} for the same values of $\lambda_{\phi}$ and $\lambda_{\psi}$.  As we show in Table \ref{tab:T3}, in this case the present value of the dark energy eos is bigger than $-1$. The eos of dark energy evolves starts from $0$ and reach to $-0.88$ at present. Also the energy density of dark energy takes over that of matter as they evolves. When we choose the positive value of $\lambda_{\psi}$, the above argument is same except the sign of $x_{\psi}$. In that case, $x_{\psi}$ decreases as it evolves with the same absolute values as the negative value of $\lambda_{\psi}$ case. Thus, the cosmological values like $\Omega_{\DE}$ and $\omega_{\DE}$ will not be changed by changing the sign of $\lambda_{\psi}$.

\begin{center}
\begin{figure}
\vspace{1.5cm}
\centerline{
\psfig{file=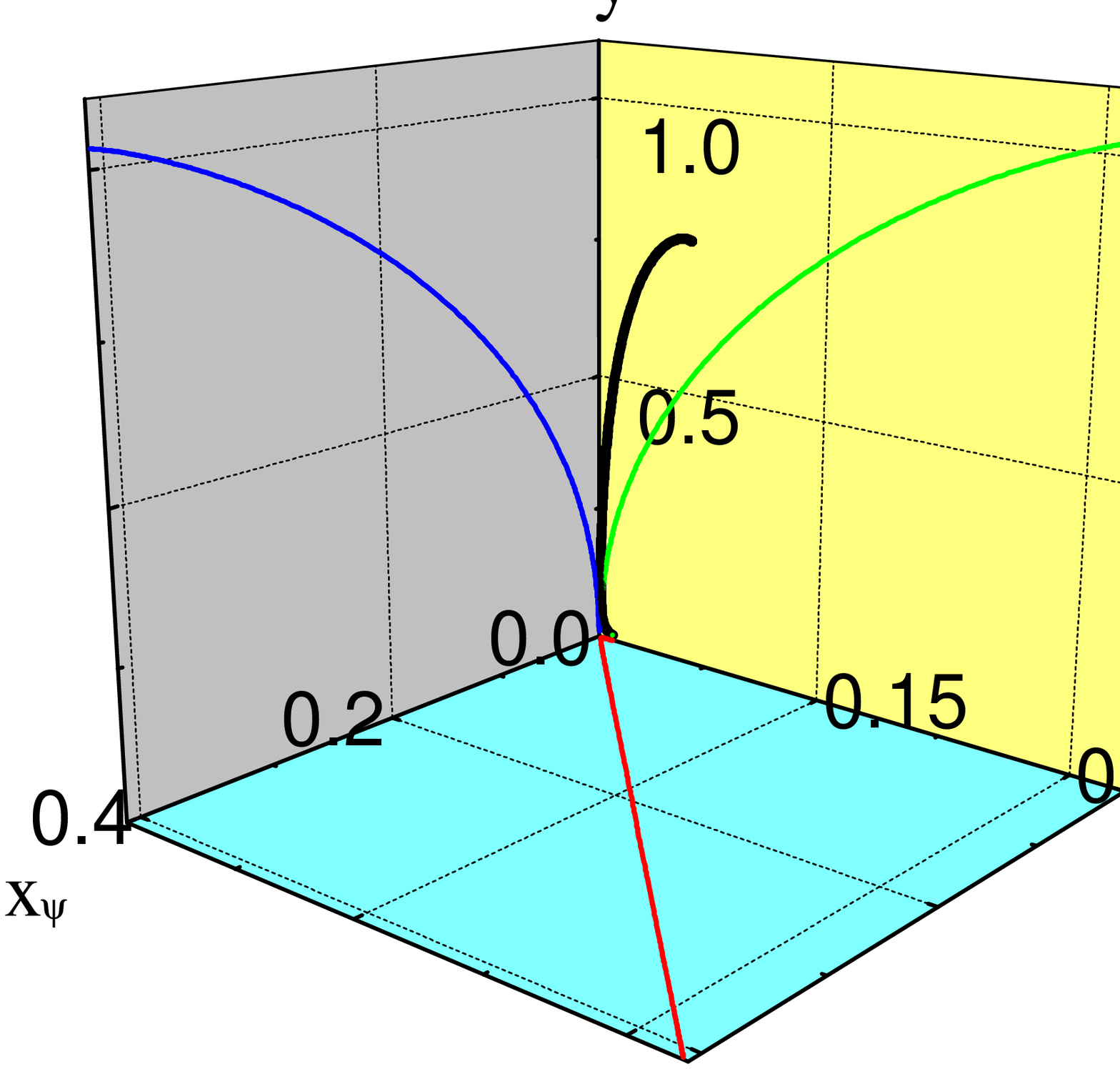, width=8cm}\psfig{file=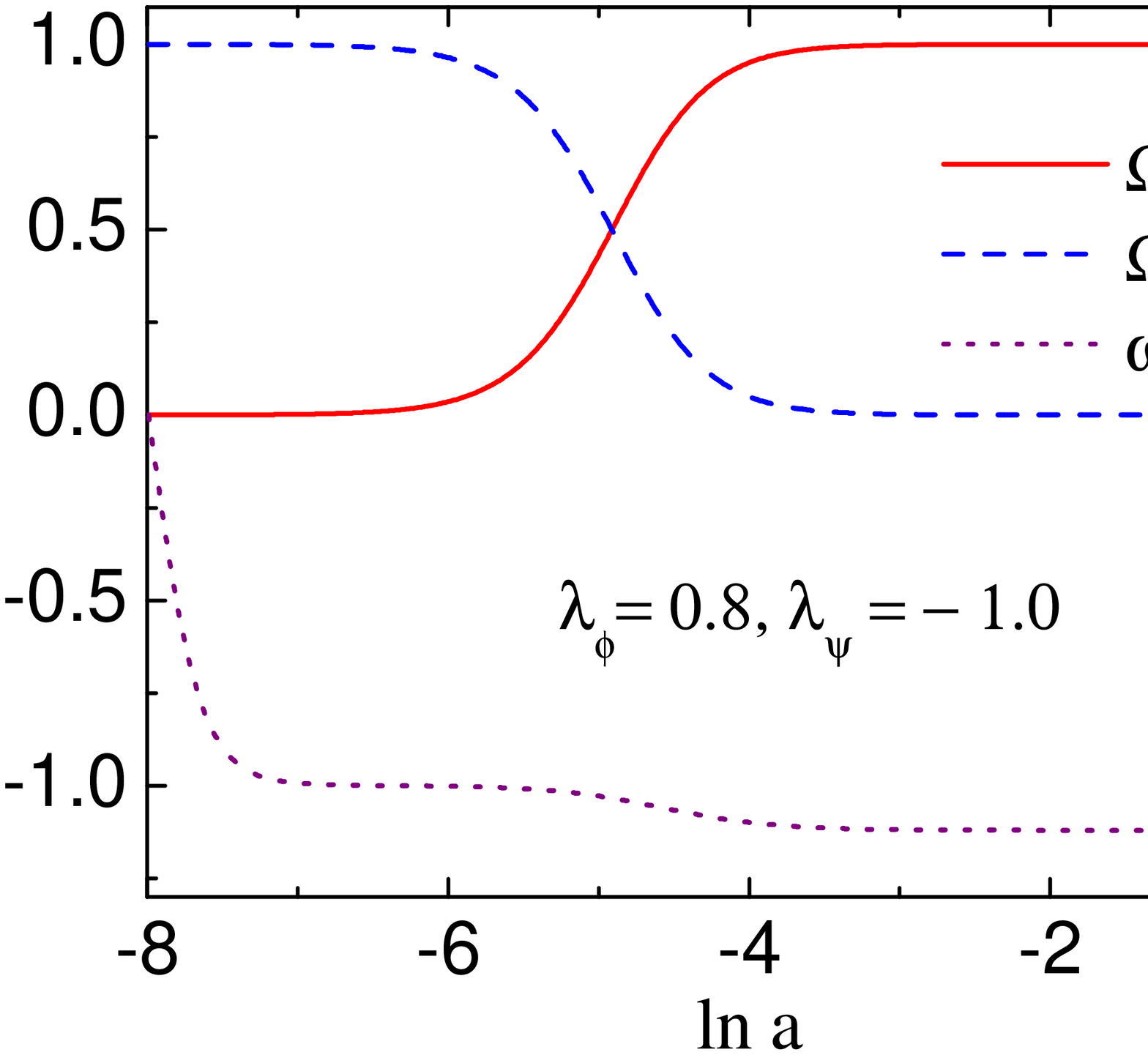, width=8cm} }
\vspace{-1.0cm}
\caption{ a) The evolution of $x_{\phi}$, $x_{\psi}$, and $y$ for $\lambda_{\phi} = 0.8$ and $\lambda_{\psi} = - 1.0$. b) The evolution of $\Omega_{\DE}$, $\Omega_{m}$, and $\omega_{\DE}$ with the same values of $\lambda_{\phi}$ and $\lambda_{\psi}$
.
} \label{fig08m10}
\end{figure}
\end{center}

Case II) $|\lambda_{\phi}| < |\lambda_{\psi}|$ :

We also show the evolutions of $x_{\phi}$, $x_{\psi}$, and $y$ for $|\lambda_{\phi}| = 0.8 < |\lambda_{\psi}| = 1.0$ in the left panel of Fig. \ref{fig08m10}. In this case, a scalar field climbs a potential slower than a LW field. Thus, $x_{\phi}$ value is smaller than $x_{\psi}$ value. Again $y$-value is increased as fields evolve but it can reach to bigger than $1$ at present. This happens because in this case $x_{\phi} < x_{\psi}$  and in order to get the $\Omega_{\DE} \leq 1$ we should have $y > 1$. In the $x_{\phi}$-$x_{\psi}$ plane, we see that both $x_{\phi}$ and $x_{\psi}$ increase from $0$ to $0.33$ and from $0$ to $0.41$, respectively. Also from both $x_{\phi}$-$y$ and $x_{\psi}$-$y$ planes, we see that $y$ starts from $0$ and reaches to $1.03$. The evolutions of $\Omega_{\DE}$, $\Omega_{m}$, and $\omega_{\DE}$ are indicated in the right panel of that figure for the same values of $\lambda_{\phi}$ and $\lambda_{\psi}$.   In this case, present value of dark energy eos is smaller than $-1$ as we can find in Table \ref{tab:T3}. The dark energy eos evolves from $0$ and reaches to $-1.12$ at present. Whenever, the condition $|\lambda_{\phi}| < |\lambda_{\psi}|$ is satisfied, we have phantom attractors at late time.

In both Fig \ref{fig10m08} and \ref{fig08m10}, we show the evolution of a scalar field and a LW field in the specific values of $\lambda_{\phi}$ and $\lambda_{\psi}$. However, the present energy density of dark energy component is too big in both case ($\Omega_{\DE}^{(0)} \simeq 1$) compared to the observed value ($\Omega_{\DE}^{(0)} \simeq 0.73$). We can see this situation from equations (\ref{OmegaDEint}) and (\ref{omegaDEint3}). In order to get the dark energy eos close to $-1$ we obtain $x_{\phi}^2 \simeq x_{\psi}^2$ from the equation (\ref{omegaDEint3}). If we use this condition to the equation (\ref{OmegaDEint}), we get $\Omega_{\DE} = y^2$. We also have $V \simeq V_{\phi-\psi}^{0}$ from the equation (\ref{Vphipsi}) because we have $\lambda_{\phi} \phi = - \lambda_{\psi} \psi$ in this case if $\phi$ and $\psi$ started from same initial values. Thus, $\Omega_{\DE} \simeq 1$ in this case and can not be matched to the observed value $\Omega_{\DE} \simeq 0.73$.

\begin{center}
\begin{figure}
\vspace{1.5cm}
\centerline{
\psfig{file=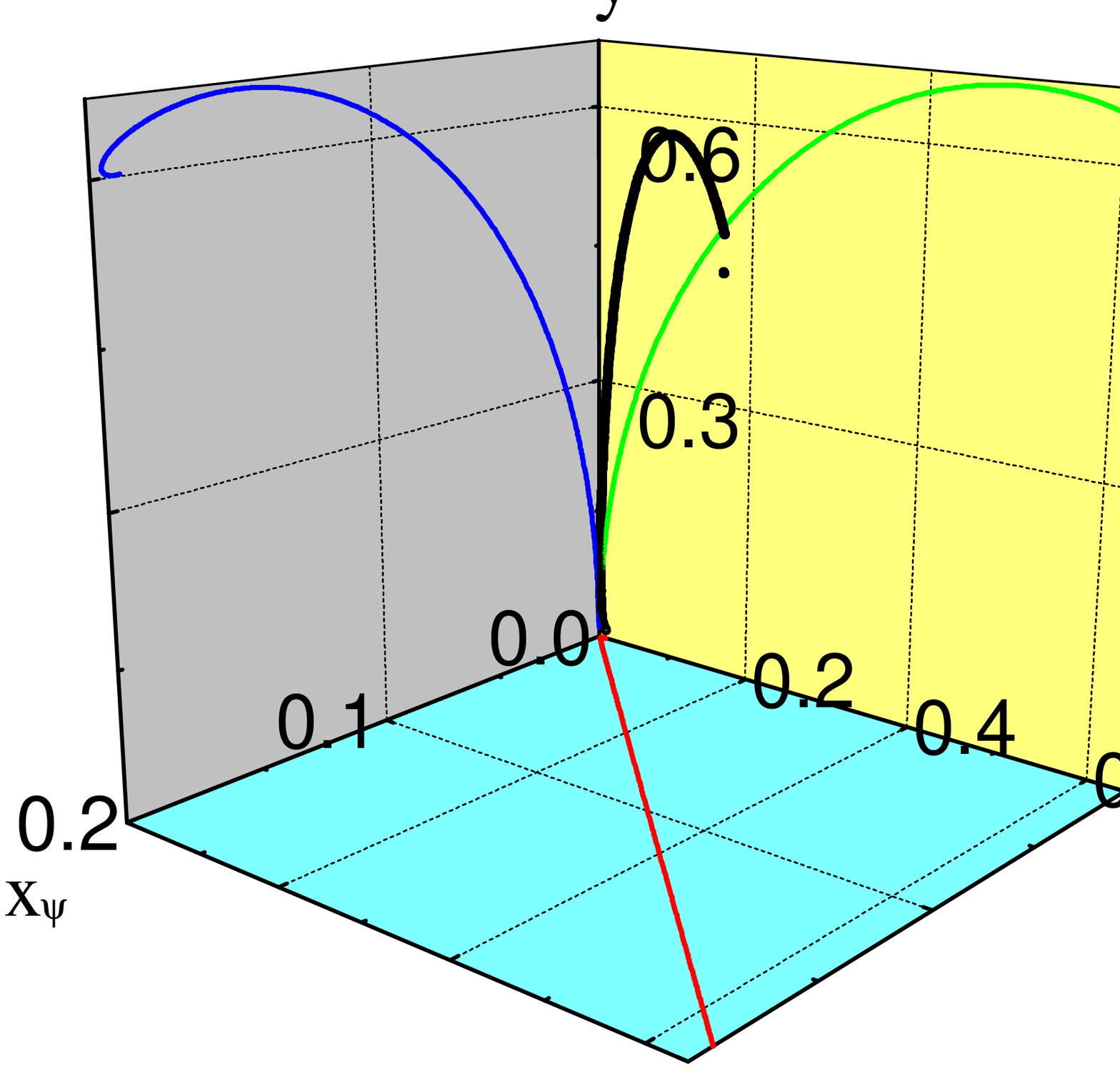, width=8cm}\psfig{file=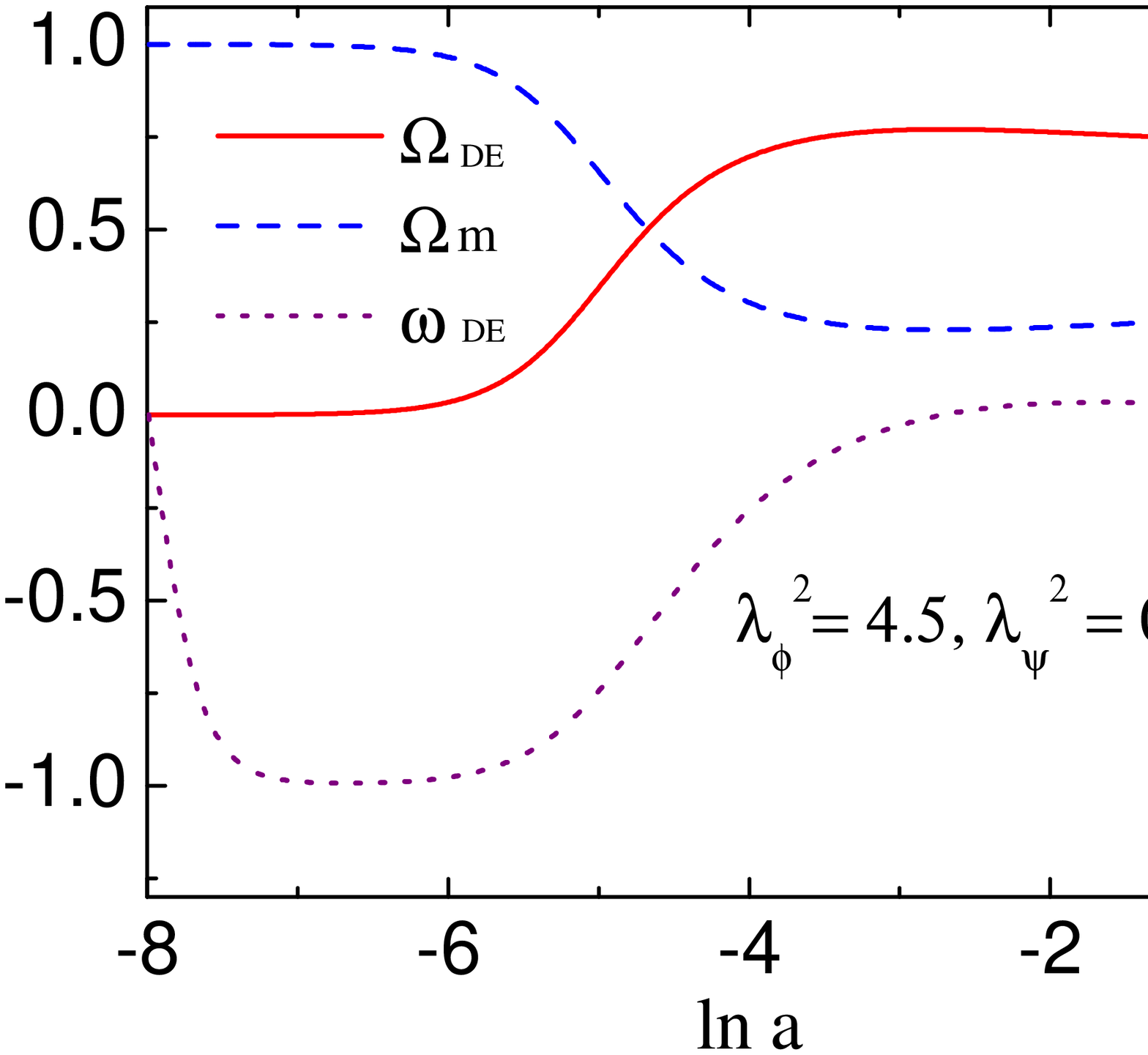, width=8cm} }
\vspace{-1.5cm}
\caption{ a) The evolution of $x_{\phi}$, $x_{\psi}$, and $y$ for $\lambda_{\phi} = \sqrt{4.5}$ and $\lambda_{\psi} = - \sqrt{0.4}$. b) The evolution of $\Omega_{\DE}$, $\Omega_{m}$, and $\omega_{\DE}$ for the same values of $\lambda_{\phi}$ and $\lambda_{\psi}$
.} \label{fig45}
\end{figure}
\end{center}

We need to tune the parameters to obtain the suitable current $\Omega_{\DE}$ value. In order to get $\Omega_{\DE}^{(0)} < 1$, we should have $\lambda_{\phi} > \lambda_{\psi}$ as shown in the above. As an example we can have $\Omega_{\DE} = 0.73$ when we choose $\lambda_{\phi} = \sqrt{4.5}$ and $\lambda_{\psi} = \sqrt{0.4}$ as shown in Fig \ref{fig45}. This choice of values satisfies the existence of dark energy-fluid dominated solution as shown in Table \ref{tab:T1int}. Nevertheless, this case follows the matter tracking attractor and the eos of dark energy converges to $0$. Thus, this case will not be able to derive the late time accelerating universe even though it dominates the late time energy component.

\section{Evolution with Self Interaction}
\setcounter{equation}{0}

\begin{center}
\begin{figure}
\vspace{1.5cm}
\centerline{
\psfig{file=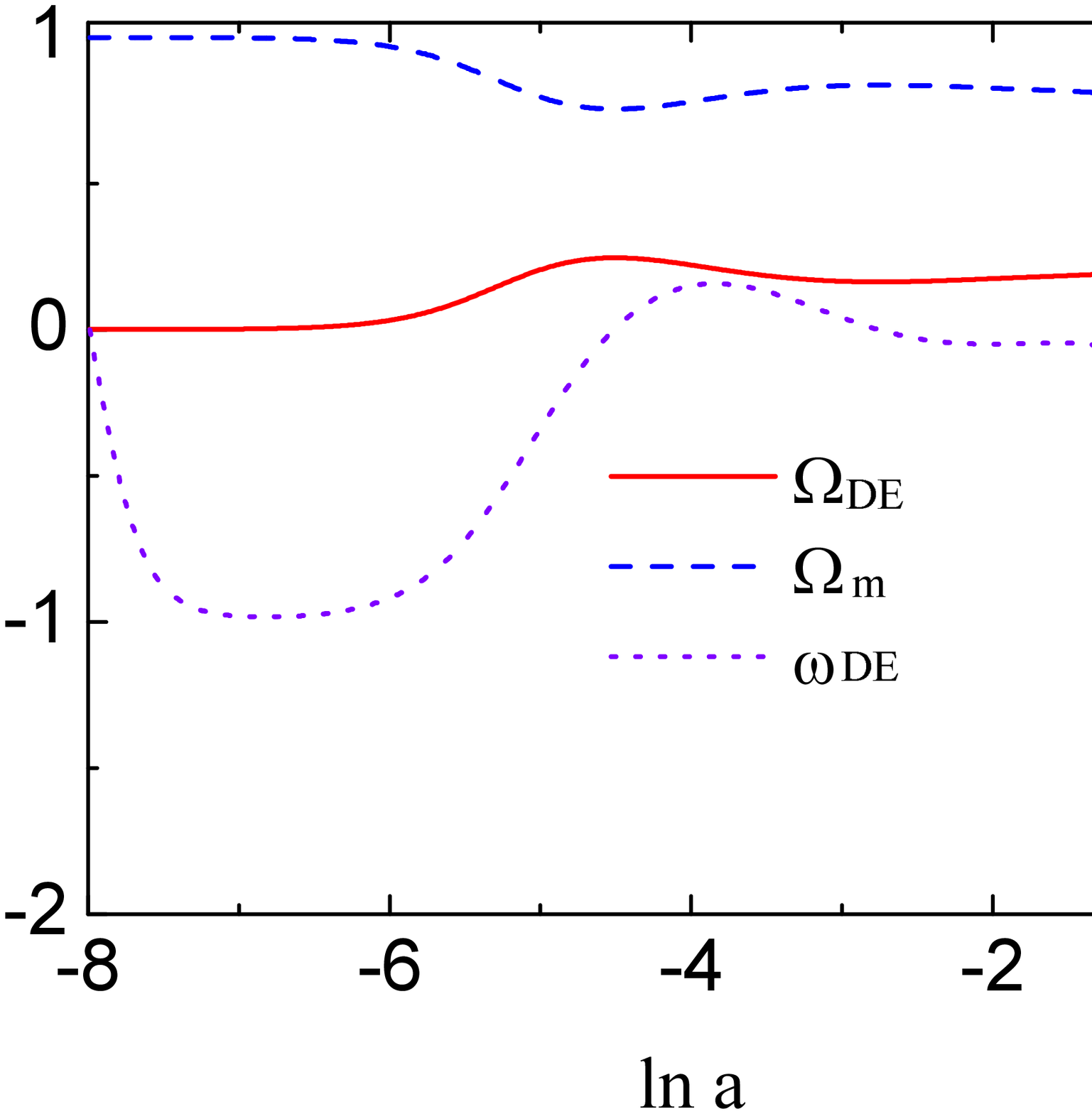, width=8cm}\psfig{file=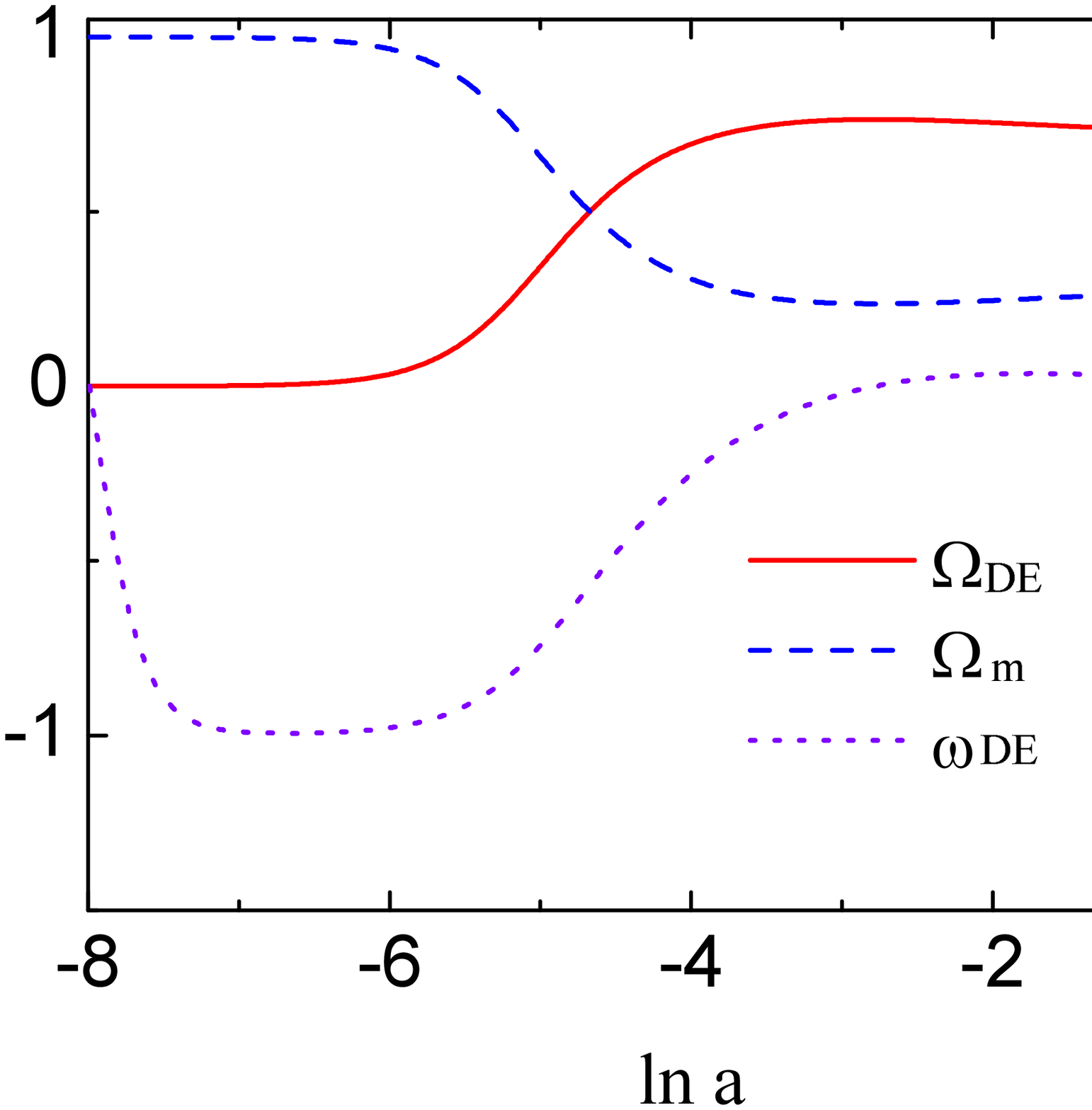, width=8cm} }
\vspace{-1.5cm}
\caption{ a) The evolution of $\Omega_{\DE}$, $\Omega_{m}$, and $\omega_{\DE}$ for $\lambda_{\phi} = 4.1$, $\lambda_{\psi} = - 0.4$, and $M = 10^{-4.5} H_{0}$. b) The evolution of same quantities for $\lambda_{\phi} = 2.1$, $\lambda_{\psi} = - 0.1$, and $M = 10^{-5} H_{0}$.} \label{fig4}
\end{figure}
\end{center}

In this section, we consider the model with including the self interaction term of LW field, $M^2 \psi^2 /2$ as in the equation (\ref{Lpsiphi}). As in the previous section, we introduce new dimensionless variables similar to equations (\ref{xyphiint} - \ref{zint}) including the self interaction term

\ba  x_{\phi} &=& \fr{\kappa \phi'}{\sqrt{6}} \, , \hspace{0.2in} y = \fr{\kappa \sqrt{V_{\rm{int}}(\phi,\psi)}}{\sqrt{3} H} \, , \label{xyphisint} \\ x_{\psi} &=& \fr{\kappa \psi'}{\sqrt{6}} \, , \hspace{0.2in} y_{\psi} = \fr{\kappa M \psi}{\sqrt{6} H} \, , \label{xypsisint} \\ z &=& \fr{\kappa \sqrt{\rho_{\gamma}}}{\sqrt{3} H} \, . \label{zsint} \ea
We can find the evolution equations of the above variables (\ref{xyphisint} - \ref{zsint}) with using the interaction term as given before (\ref{Vphipsi}) 

\ba x_{\phi}' &=& - \Biggl( \fr{H'}{H} + 3 \Biggr) x_{\phi} + \fr{\sqrt{6}}{2} \lambda_{\phi} y^2 \, ,\label{xphipsint} \\ y' &=& - y \Biggl( \fr{H'}{H} + \fr{\sqrt{6}}{2} ( \lambda_{\phi} x_{\phi} + \lambda_{\psi} x_{\psi}) \Biggr) \, ,\label{ypsint} \\ x_{\psi}' &=& - \Biggl( \fr{H'}{H} + 3 \Biggr) x_{\psi} + \fr{M}{H} y_{\psi} - \fr{\sqrt{6}}{2} \lambda_{\psi} y^2 \, , \label{xpsipint} \\ y_{\psi}' &=& \fr{M}{H} x_{\psi} - \fr{H'}{H} y_{\psi} \, . \label{ypsipint} \ea
Compared to the without self interaction case, we are not able to have the critical points in this system. However, we can still investigate the cosmological evolution of this system. 

In the left panel of Fig \ref{fig4}, we show the evolution of $\Omega_{\DE}$, $\Omega_{m}$, and $\omega_{\DE}$ when we choose that $\lambda_{\phi} = 4.1$, $\lambda_{\psi} = - 0.4$, and $M = 10^{-4.5} H_{0}$. In this case, we have $\Omega_{\DE}^{(0)} = 0.73$ and $\omega_{\DE}^{(0)} = -1.83$. Thus, this case can be  viable for the current observations. In the right panel of this figure, we show the cosmological evolutions of same quantities for $\lambda_{\phi} = 2.1$, $\lambda_{\psi} = - 0.1$, and $M = 10^{-5} H_{0}$. With this specific choice of parameters we have $\Omega_{\DE}^{(0)} = 0.73$ and $\omega_{\DE}^{(0)} = -0.01$. This is not be able to produce the late time acceleration universe even though the dark energy component becomes a dominate component in the universe at late time. Also as we can see in the figure, the energy density of the dark energy dominates the universe too early.

\section{Conclusion}
\setcounter{equation}{0}

We show a dark energy candidate model from LW theory. With two scalar fields we can have the late time equation of state of the dark energy across $-1$ which might be suitable for the current observations. Even though the structure of the theory is similar to the so called ``quintom'' model, we can not avoid the interaction between two scalar fields when we start from LW theory. Also the signs of slope of potential are opposite to each other. Interestingly, there exist the late time tracking attractor solutions for the specific choice of the slopes of a potential. However, this simple exponential potential models can not give the viable models which derive the current late time accelerating universe. We also investigate the model with including the self interaction term. In this case we can produce the suitable cosmological evolution of observable quantities. However, we are not able to find the stable solution in this case and we also suffer from the fine tuning for the mass of LW field. 

\section*{Acknowledgments}

We thank Y-Y. Keum for useful discussion.

\end{document}